\documentclass[12pt]{article}

\usepackage{amsmath}
\usepackage{amssymb}
\usepackage{indentfirst}
\usepackage{amssymb}
\usepackage{amsfonts}
\usepackage{amscd}
\usepackage{amsbsy}
\usepackage{amsthm}
\usepackage{latexsym}
\usepackage{graphicx,color} 
\usepackage[dvipsnames]{xcolor}
\usepackage[colorlinks]{hyperref}
\hypersetup{linkcolor=blue,citecolor=blue,urlcolor=blue}

\def\nn{\nonumber}       
\def\beq{\begin{eqnarray}}
\def\eeq{\end{eqnarray}}

\def\ln{\,\mbox{ln}\,}

\def\Tr{\,\mbox{Tr}\,}

\def\al{\alpha}
\def\be{\beta}

\def\ga{\gamma}

\def\ka{\kappa}
\def\la{\lambda}

\def\pa{\partial}

\def\si{\sigma}

\def\La{\Lambda}

\usepackage[symbol]{footmisc} 
\usepackage{float} 
\usepackage{cite}  
\usepackage{titlesec}

\usepackage{ulem} 

\usepackage{pgfplots}
\usepackage{tikz}
\pgfplotsset{width=7cm,compat=1.18}
\usepackage{setspace} 
\usepackage{type1cm} 
\usepackage{float} 

\begin{document}
	
\begin{center}
	\renewcommand*{\thefootnote}{\fnsymbol{footnote}}
	
	{\Large
	Possibility of Spontaneous Symmetry Breaking in the
	 \\ 
	 Nambu-Jona-Lasinio model  with torsion 
		 }
	\vskip 6mm
	
	{\bf P\'ublio Rwany B. R. do Vale}
	\footnote{E-mail address: \ publio.vale@gmail.com}
	\vskip 6mm
	
	Departamento de F\'{\i}sica, ICE, Universidade Federal de Juiz de Fora,
	\\
	36036-900, Juiz de Fora, Minas Gerais, Brazil
\end{center}

\vskip 2mm
\vskip 2mm


\begin{abstract}
	
	\noindent
	We discuss the Nambu-Jona-Lasinio (NJL) model in curved spacetime with torsion in the leading order of the $1/N$  expansion. The effective potential of the scalar-torsion sector is calculated using the new technique based on the nonlocal part of anomaly-induced action which was recently found to produce the effective potential in the low-energy limit. The spontaneous symmetry breaking caused by the change of the torsion term is found, confirming the statements known from the existing literature. Furthermore, the gap equation is calculated as a function of curvature and torsion. Finally, the behavior of the effective four-fermion coupling constant as a function of the torsion is discussed.
	\vskip 3mm
	
	
\end{abstract}

\setcounter{footnote}{0} 
\renewcommand*{\thefootnote}{\arabic{footnote}} 

\section{Introduction}
 Élie Cartan, in 1922, proposed the Einstein-Cartan theory (ECT) of gravity,
 a modification of the general relativity theory (GR). That change introduced by the ECT allowed the existence of torsion in spacetime in addition to curvature. A review of the fundamentals of the ECT can be found in e.g., \cite{Hehl (73),Hehl (76),Trautman,Shap-Tors (02)}.
 The ECT, also called Einstein-Cartan-Sciama-Kibble (ECSK), became
 an important tool due to its range of applications in particle physics
 and cosmology \cite{Barut (87),Freidal (05),Ribas (05),Taveras (08),Gomez (09),Kremer (10),Diakonov (11),Peixoto (12),Magueijo (13),Khriplovich (13)}.
 
In \cite{Perez (06)} describes a new theory of gravitation that combines the ECT minimally coupled to fermions in addition to the Holst action, showing that this gravitational theory is equivalent to the Einstein-Hilbert (EH) action with four-fermion interaction term and Dirac action.
 
 Recently, was demonstrated in \cite{Shaposhnikov (20)} that the
 four-fermion interactions may not only mediate the production of feebly
 interacting fermions right after inflation, but they can play the
 role of dark matter if such fermions are singlets concerning the Standard
 Model (SM) gauge group. This possibility is explored in \cite{Shaposhnikovv (21)},
 where it is demonstrated that the four-fermion interactions
 in a wide range of fermion masses could generate the observed dark
 matter abundance.
 
 The Nambu-Jona-Lasinio (NJL) model \cite{NJL(61)} is an effective four-fermion theory consistent with the SM gauge group and compatible with some interpretations of the fermionic axial currents contained in the EH action with  Dirac action and four-fermion interaction, see e.g., \cite{Klev(92),Buballa(05)} for review.  
 
 Detailed studies of the NJL model have already been performed in
 curved spacetime, e.g., first in \cite{Hill (92)}, in the study of inflation \cite{Marc}, and phase transition \cite{Eliz(94),Inag(93),Inagaki (97)},
 and also in the presence of an external torsion for the renormalization group analysis  \cite{Shap(94)}.
 In \cite{Inag(93)} carried out a systematic study of the phase structure of the NJL model in external gravity with the idea of investigating how the phase transition occurs in the circumstances of the early universe. That is, verify as mechanics
 of spontaneous symmetry breaking take place in the NJL model in external
 gravity.

 In summary, we can say that the NJL model can be seen as a prototype model, as it allows investigations in different scenarios. This is because the NJL model is a simple four-fermion interaction model, and its quantum field theory analysis that takes loop corrections into account would not be complicated to achieve, see e.g., \cite{Coleman (73)}. As the study on torsion is on the rise, since we can assume that torsion plays an important role in the formation of dark matter  \cite{Shaposhnikov (20),Shaposhnikovv (21)}, and recently it was demonstrated the possibility of symmetry breaking in the torsion sector for a fermionic action \cite{Gui-Shap(22)}. Inspired by \cite{Inag(93)}, we want to understand how the spontaneous symmetry breaking mechanism occurs in the NJL model under the influence of external torsion.
 
 In this paper, the subject of study is the gap equation with external background composed by metric and torsion fields. On the other hand, in the literature there are some papers that discuss dynamical torsion in other theories, see \cite{Shap-Tors (02),Helayel (00),Peixoto (07)} and the recent papers \cite{Martini (23),Martini (24)}.
 
 The paper is organized as follows. In Sec. 2 we repeat the considerations of the papers \cite{Shap-Tors (02)}, to have all necessary information for further consideration. Therefore, the equivalence between the ECT in conjunction with the Holst action, when coupled to fermions, and the EH action with the Dirac action, along with a four-fermion interaction term, will be established, as asserted by \cite{Perez (06)}. Subsequently in  Sec. 3, we briefly review the NJL model for gravity with torsion and find the classical effective action through the auxiliary fields method, i.e., using the bosonization procedure. Sec. 4 describes the calculation of the conformal anomaly in the fermion case with scalar field. The conformal anomaly \cite{Duff (74),Duff (76),Duff (77),Duff (94)} is used to find the nonlocal part solution of anomaly-induced action \cite{Riegert (84),Fradkin (84)}(see for example \cite{Mottola (92),Shapiro (08)} for review and further reference). In Sec. 5 using the nonlocal form of anomaly-induced action we can describe the low-energy limit in the effective action in the metric-torsion theory. Sec. 6 the effective potential and the gap equation of the NJL model with the scalar-torsion sector are calculated and subsequently analyzed. In Sec. 7, we draw our conclusions.

\section{ECT with Holst term and Dirac action}
In this section we closely follow the papers \cite{Shap-Tors (02),Perez (06),Shaposhnikov (20),Shaposhnikovv (21)}, to have all necessary information for further considerations.
One motivation for exploring the ECT coupled to fermions with the Holst term lies in the emergence of the four-fermion interaction term appearing in the action after solving for torsion, as we can show below through the following action
\begin{equation}\label{eq:1.1}
	S=
		\dfrac{1}{\kappa^2}\int d^4 x\,\sqrt{-g}\left[-\tilde{R}+\dfrac{1}{2\beta}\epsilon_{\al\be\mu\nu}\tilde{R}^{\al\be\mu\nu}\right]+\dfrac{i}{2}\int d^4x\,\sqrt{-g}\left[\bar{\psi}\gamma^{\mu}\tilde{\nabla}_\mu \psi-\tilde{\nabla}_\mu\bar{\psi}\gamma^\mu\psi\right],
\end{equation} 
where the constant $\kappa^2=16\pi G$, being $G$ the Newton constant, $g$ is the determinant of the spacetime metric $g_{\mu\nu}$ and $\tilde{R}$ is the scalar curvature with torsion, which is given by
\begin{equation}\label{eq:1.1-2}
	\tilde{R}=R-2\nabla_{\alpha}T^{\alpha}-\dfrac{2}{3}T_{\alpha}T^{\alpha}+\dfrac{1}{24}S_{\alpha}S^{\alpha}+\dfrac{1}{2}q_{\alpha\beta\gamma}q^{\alpha\beta\gamma},
\end{equation}
with $R$ being the Ricci scalar in GR,
and the torsion being divided into three irreducible components, which are the trace vector $T_\be=T^{\al}{}_{\be\al}$ that is antisymetric in the last two indices, the axial vector (or pseudotrace) $S^\nu=\epsilon^{\al\be\mu\nu}T_{\al\be\mu}$, and the tensor $q^{\al}{}_{\be\ga}$ which satisfies two conditions $q^{\al}{}_{\be\al}=0$ and $\epsilon^{\al\be\mu\nu}q_{\al\be\mu}=0$.

The second term of right-hand side (\textit{r.h.s.}) of equation (\ref{eq:1.1}) is the Holst term
\begin{equation} \label{eq:1.1-3}
	\dfrac{1}{2\beta}\epsilon_{\al\be\mu\nu}\tilde{R}^{\al\be\mu\nu}
		=\dfrac{1}{2\beta}\left[
		-\nabla_{\mu}S^\mu-\dfrac{2}{3}T^\mu S_\mu+\dfrac{1}{2}\epsilon_{\al\be\mu\nu}q^{\la\al\be}q_{\la}{}^{\mu\nu}\right],
\end{equation}
being $\tilde{R}^{\alpha\beta\mu\nu}$  the curvature tensor in the space-time with torsion, $\beta$ the Barbero-Immirzi parameter and $\epsilon_{\al\be\mu\nu}$ the Levi-Civita symbol. 

The Dirac action with torsion is the last term of \textit{r.h.s.} of equation (\ref{eq:1.1}), where the term $\gamma^\mu=e_{a}{}^{\mu}\gamma^{a}$ is the Dirac matrix in curved spacetime, the quantities $e_{a}{}^{\mu}$ is tetrad field, and  the covariant derivative in the curved spacetime with torsion of the fermion field is written as follows 
\begin{equation} \label{eq:1.1-4}
	\tilde{\nabla}_{\mu}\psi=\partial_\mu\psi
		+\dfrac{i}{2}\tilde{\omega}_\mu{}^{ab}\sigma_{ab}\psi,
		\quad
		\textrm{and}
		\quad
		\tilde{\nabla}_\mu\bar{\psi}=\partial_\mu\bar{\psi}
		-\dfrac{i}{2}\bar\psi\,\tilde{\omega}_\mu{}^{ab}\sigma_{ab}\,,
\end{equation}
where $\sigma_{ab}=\frac{i}{2}\left(\gamma_{a}\gamma_{b}-\gamma_{b}\gamma_{a}\right)$. The spinor connection with torsion and without torsion is given by 
\begin{align} \label{eq:1.1-6}
	\tilde{\omega}_{\mu}{}^{ab} =
		\omega_{\mu}{}^{ab}+\dfrac{1}{4}
		K^\alpha{}_{\lambda\mu}
		\left(e^{\la a}e^{b}{}_{\al}-e^{\la b}e^{a}{}_{\al}\right) 
\quad\textrm{and}\quad
	K^{\al}{}_{\la\mu}=
	\dfrac{1}{2}
	\left(T^{\al}{}_{\la\mu}-T_{\la}{}^{\al}{}_{\mu}+T_{\mu}{}^{\al}{}_{\la}\right),
\end{align}
where $\omega_{\mu}{}^{ab}$ is spinor connection without torsion. The second equation in (\ref{eq:1.1-6}) is contorsion tensor that is antisymmetric in the first two indices.
It is important to highlight that the indices $a,b,c,\ldots=0,1,2,3$ are internal Lorentz indices while the Greek indices $\mu,\nu,\alpha,\ldots=0,1,2,3$ are spacetime indices.

Replacing the previous equations (\ref{eq:1.1-2}), (\ref{eq:1.1-3}) and (\ref{eq:1.1-4})  into (\ref{eq:1.1}), and subsequently after some calculations, we can rewrite equation (\ref{eq:1.1}) in the form
\begin{align} \label{1.6}
		S=&\dfrac{1}{\ka^2}\int d^{4}x
		\left[
		-R+\dfrac{2}{3}T_{\al}T^{\al}-\dfrac{1}{24}S_{\al}S^{\al}-\dfrac{1}{3\beta}T_{\al}S^{\al}	\right] 
		\nonumber
		\\
		&
		+\dfrac{1}{8}\int d^4x \sqrt{-g}S_\al J^\al
		+\dfrac{i}{2}\int d^4x\sqrt{-g}
		\left[\bar{\psi}\ga^\mu\nabla_{\mu}\psi
		-\nabla_{\mu}\bar{\psi}\ga^\mu\psi\right],
\end{align}
where $J_\mu=\bar{\psi}\ga_5\ga_{\mu}\psi$ represents the fermionic axial current. At this point, we can make a few observations. The first term on the \textit{r.h.s.} of the equation above is related to the ECT with the Holst term, that is, to the first integral on the \textit{r.h.s.} of the equation (\ref{eq:1.1}). In contrast, the last two integrals in the previous equation, in turn, correspond to the spinor action, equivalent to the second integral on the \textit{r.h.s.} of equation (\ref{eq:1.1}).

When considering only the spinor action in (\ref{1.6}), we observe that only the axial component of torsion couples with fermions, while the other components decoupled. However, when analyzing equation (\ref{1.6}) in its entirety, we can determine the axial vector $S^\al$ and the trace vector $T^\al$ through their respective equations of motion, which are $S^\al=16\kappa^2\xi J^\al$ and $T^\al=\left(1/4\be\right)S^\al$, with $\xi$ being a kind of coupling constant defined by 
\begin{align}
	\xi=\dfrac{3}{32}\left(\dfrac{\beta^2}{1+\beta^2}\right)
	\nonumber.
\end{align} 
Replacing the equations of motion into (\ref{1.6}), we obtain
\begin{equation}
	S=
		\dfrac{i}{2}\int d^4x\sqrt{-g}
		\left[\bar{\psi}\ga^\mu\nabla_{\mu}\psi
		-\nabla_{\mu}\bar{\psi}\ga^\mu\psi\right]
		+\dfrac{1}{\ka^2}\int d^{4}x\sqrt{-g}\left[-R+\xi\ka^{4}J_{\mu}^{2}\right]
		\, 
		.\label{EH+4f+D}
\end{equation}
In this expression, the action contains the  EH action and the fermion kinetic term, with the last term on the \textit{r.h.s.} describing the four-fermion interaction by antisymmetric torsion, see e.g., \cite{Shap-Tors (02),Perez (06)} for details. In \cite{Shaposhnikov (20), Shaposhnikovv (21)}, it was demonstrated that the fermionic current generated by torsion in the ECT, i.e., the four-fermion interaction in equation (\ref{EH+4f+D}), could be a possible candidate for dark matter. Given that the fermionic axial current can be seen as a potential candidate for dark matter, it is interesting to explore how the non-dynamical
torsion, could impact the phase transition of the NJL model.
	
\section{Description and effective potential for the NJL model}

Let us start with the NJL model 
in an external gravitational field with torsion  in
the $1/N$ expansion (see, e.g., \cite{Witten (79),Sazdjian (15)} for review and \cite{Rosenstein(941)} in four-fermion interaction models) described by the Lagrangian
\begin{equation}
	S_{\textrm{NJL}}=\int d^{4}x\sqrt{-g}\left\{ i\bar{\psi}\gamma^{\mu}\left(\nabla_{\mu}-i\eta\gamma_{5}S_{\mu}\right)\psi+\dfrac{\lambda}{2N}\left[\left(\bar{\psi}\psi\right)^{2}+\left(\bar{\psi}i\gamma_{5}\psi\right)^{2}\right]\right\} ,\label{eq:2.1}
\end{equation}
where $N$ is the number of fermions, the coupling constant recast
$\lambda$, and $\eta$ is the 
nonminimal coupling constant and its value is $1/8$, see, e.g., \cite{Shap-Tors (02),Shap(94),Shap (92)}.
The interaction terms are constituted by the scalar channel  $\left(\bar{\psi}\psi\right)^{2}$
and the pseudoscalar chanel $\left(\bar{\psi}i\gamma_{5}\psi\right)^{2}$. 

As discussed in the previous section, it was expected that only the axial component of torsion would couple to the fermions. Due to the absence of the Einstein-Hilbert (EH) term and the Holst term in the action (\ref{eq:2.1}), the axial vector $S_\mu$, in this context, will not be proportional to the axial current but will act as a external field.

The bosonization procedure the mean field approach consists of introducing the auxiliary fields $\rho$ and $\pi$, which are scalar and pseudoscalar fields, into the above action. The auxiliary fields  will be inserted in the form
\begin{equation}
	S\rightarrow S
	-\int d^4x\sqrt{-g}\left\{\dfrac{N}{2\lambda}\left[\rho+\dfrac{\lambda}{N}\bar{\psi}\psi\right]^{2}
	+\dfrac{N}{2\lambda}\left[\pi+\dfrac{\lambda}{N}\bar{\psi}i\gamma_5\psi\right]^{2}\right\}.
\end{equation}
The action classically equivalent to (\ref{eq:2.1}) has the form
\begin{equation}
	S_{\textrm{NJL}}=\int d{{}^4}x\sqrt{-g}\left[i\bar{\psi}\gamma^{\mu}\left(\nabla_{\mu}-i\eta\gamma_{5}S_{\mu}\right)\psi-\dfrac{N}{2\lambda}\left(\rho{{}^2}+\pi{{}^2}\right)-\bar{\psi}\left(\rho+i\gamma_{5}\pi\right)\psi\right].\label{eq:2.2}
\end{equation}
The Euler-Lagrange equation for the auxiliary fields in the previous equation has the following form,
\begin{align}
	\rho=-\dfrac{\lambda}{N}\bar{\psi}\psi
	\quad\textrm{and}\quad
	\pi=-\dfrac{\lambda}{N}\bar{\psi}i\gamma_{5}\psi ,
	\nn
\end{align}
when replaced in (\ref{eq:2.2}), produce (\ref{eq:2.1}).

Our purpose is to find the effective action by applying the integral path quantization technique. Let us start with the generating of the Green functions of the fermion field\footnote{It is also possible to integrate over the torsion field (see \cite{Martini (23),Martini (24)}), but this requires promoting torsion to be dynamical and may lead to the inconsistencies of the effective theory \cite{Peixoto (07)}.}  
\begin{equation}
	Z\left[\theta,\bar{\theta}\right]=\int D\psi D\bar{\psi}D\rho D\pi\exp\left(iS+i	\theta\bar{\psi}+i\psi	\bar{\theta}\,\right),\label{eq:2.3}
\end{equation}
where $\theta$ and 	$\bar{\theta}$ are Grassmannian sources. Integrating over
the fermion fields, and after some calculations, we obtain the following generating functional (see \cite{Klev(92)} for more details) 
\begin{equation}
	Z=\int D\rho D\pi \exp\left\{
	iNS_{\textrm{eff}}
	+\int d^4x\sqrt{-g}
	\,\bar{\theta}
	\left[\dfrac{1}{i\gamma^\mu\left(
	\nabla_\mu-i\eta \gamma_5 S_\mu	\right)-s}
	\right]
	\theta
	\right\}.
\end{equation}
The semiclassical effective action is given by
\begin{equation}
	S_{\textrm{eff}}=\int d^{4}x\sqrt{-g}\left[-\dfrac{1}{2\lambda}\left(\rho^{2}+\pi^{2}\right)\right]-i\textrm{Tr} \ln\left[i\gamma^{\mu}\left(\nabla_{\mu}-i\eta\gamma_{5}S_{\mu}\right)-	s\right], \label{eq:2.5}
\end{equation}
where $s=\rho+i\gamma_5 \pi$. 

We know that the leading term of the
effective action $\Gamma\left[\rho,\pi\right]$ is just equal to $S_{\textrm{eff}}$. As can be seen from the equation below
\begin{equation}
	\Gamma\left[\rho,\pi\right]=S_{\textrm{eff}}\left[\rho,\pi\right]+\mathcal{O}\left(1/N\right)\label{eq:2.6},
\end{equation}
	the effective action can be write as a derivative expansion, where the effective potential correspond to the zeroth order term, i.e.,
\begin{equation} \label{Geff}
		\Gamma\left[\rho,\pi\right]=
	\int d^4x\sqrt{-g}
	\left[-V_{\textrm{eff}}
	+\dfrac{1}{2}Z\left(\rho\right)g^{\mu\nu}\pa_\mu\rho\,\pa_\nu\rho
	+\dfrac{1}{2}Z\left(\pi\right)g^{\mu\nu}\pa_\mu\pi\,\pa_\nu\pi
	+\ldots\right]\, .
\end{equation}
Obviously, we can find the effective potential in equation  (\ref{Geff}), which can be written as
\begin{equation}
	V_{\textrm{eff}}=V_{\textrm{flat}}+V_{R}+V_{S^{2}}.
\end{equation}
Here effective potential is naturally separated into three different
parts, namely, the potential in flat spacetime $V_{\textrm{flat}}$,
the potential in curved spacetime without torsion $V_{R}$ and the contribution of
the torsion term $V_{S^{2}}$. Let is note that the terms $V_{\textrm{flar}}$ and
$V_{R}$ already are known and were discussed in \cite{Marc,Eliz(94),Inag(93)}. Our purpose is to find the contribution of torsion in the effective
potential. Fortunately, we can find $V_{S^{2}}$ using the efficient and technically simple technique based on the nonlocal part of anomaly-induced action, as already demonstrated in an analogous theory in \cite{Gui-Shap(22),Shap et All (22)}.

\section{Anomaly-induced effective action }

Briefly, we can say that the anomaly-induced effective action is a simple construction aiming to describe loop corrections in the semiclassical approach.
The anomaly and induced action are almost equivalent to the local version of the usual renormalization group approach \cite{Coleman (73)}, but provide a more detailed output, as explained, e.g. in \cite{Shapiro (08)}.
Here we follow this approach as it is the most economic one, and gives the same result as other methods.
In this section, we will use the conformal anomaly to find the covariant nonlocal solution of anomaly-induced action and, subsequently, in the next section, extract from it the effective potential. 

Let us consider a model with Dirac spinor fields and scalar field that is also invariant under the local conformal transformations\footnote{
		The conformal transformation of the axial vector part of torsion does not follow the conformal transformation for the metric and torsion in the ECT. In this particular theory,  there is a proportionality between $S_{\mu}$ and the axial fermionic current, which allows the following conformal transformation, as 
	\begin{equation}
		S^{\mu}=e^{-4\sigma}\bar{S}^{\mu}
			\quad\textrm{and}\quad
			S_{\nu}=e^{-2\sigma}\bar{S}_{\nu}. \nn
	\end{equation}
	However, this is not a universal rule. The action of torsion,  in our case, has additional terms and the connection is more complicated. In fact, we have the freedom to choose the conformal transformation as we like and it is taken to be such that the classical action possesses conformal invariance and we can use trace anomaly to evaluate quantum corrections to the potential of scalar and torsion, see \cite{Shap-Tors (02), Buchbinder(85)} for details.} 
\begin{align} \label{C-T-R}
	g_{\mu\nu}=\bar{g}_{\mu\nu}e^{2\sigma},\quad
	S^{2}=\bar{S}^{2}e^{-2\sigma},\quad
	S_{\mu\nu}^{2}=\bar{S}_{\mu\nu}^{2}e^{-4\si},
	\quad \textrm{and} \quad
	\rho=\bar{\rho}e^{-\sigma},
\end{align}
whose bilinear form is given by
\begin{equation}
	\hat{H}=i\gamma^{\mu}\left(\nabla_{\mu}-i\eta\gamma^{5}S_{\mu}\right)-\rho.\label{eq:3.1}
\end{equation}
To evaluate the one-loop corrections  $\bar{\Gamma}{}^{(1)}=-i\Tr\ln\hat{H}$, we use the Schwinger-DeWitt technique \cite{DeWitt(65), Bar-Vilk(85)} to find the divergences and, later, the expression for the conformal anomaly, i.e., the anomalous trace. In the particular case of antisymmetric torsion and real scalar, the anomaly has the form
\begin{align} \label{anomaly}
	\left\langle T\right\rangle   
	=-\left\{ wC^{2}+bE_{4}+c\square R	 
	-\beta_{1}\rho^{2}S^{2}
	-\dfrac{1}{4}\beta_{2}S_{\mu\nu}^{2}
	\right\} 
	+\textrm{surface terms},
\end{align}
compared to the similar expression in \cite{Gui-Shap(22)}, we removed irrelevant surface terms as well as purely scalar terms, and subsequently added contributions from the scalar-torsion sector. Regardless of these changes, the consequent considerations are essentially uncharged.

 Since $C^2=R^{2}_{\mu\nu\al\be}-2R^{2}_{\al\be}+\frac{1}{3}R^{2}$ is the square of the Weyl tensor in four-dimensional spacetime, being $R_{\mu\nu\al\beta}$ the Riemann curvature tensor and $R_{\mu\nu}$ is Ricci curvature tensor. The term $E_4=R^{2}_{\mu\nu\al\be}-4R^{2}_{\mu\nu}+R^2$,
is the integrand of the Gauss-Bonnet topological term and 
	$S_{\mu\nu}=\nabla_{\mu}S_{\nu}-\nabla_{\nu}S_{\mu}$. 

The one-loop $\beta$-functions $w,b$ and $c$ in the vacuum sector 
can be found in \cite{Shap-Tors (02), Gui-Shap(22), Bir-Dav(82), Buch-Shap(21), Buchbinder(85)}, and the coefficients $\beta_{1,2}$ are given by
\begin{equation}
		\beta_{1}
		=-\dfrac{8}{\left(4\pi\right)^{2}}N\eta^{2}
		\quad
		\textrm{and}
		\quad
		\beta_{2}
		=\dfrac{8}{3\left(4\pi\right)^{2}}N\eta^{2},\label{eq:3.7}
\end{equation}
where $N$ the number of the copies of the fermions.

The anomaly-induced action can be determined through the solution of equation 
\begin{equation} \label{Tra-IndAc}
		-\dfrac{2}{\sqrt{-g}}g_{\mu\nu}\dfrac{\delta\Gamma_{\textrm{ind}}}{\delta g_{\mu\nu}}
	+
		\dfrac{1}{\sqrt{-g}}\rho \dfrac{\delta \Gamma_{\textrm{ind}}}{\delta \rho}
	=
		-\dfrac{1}{\sqrt{-\bar{g}}}e^{-4\sigma}
		\left.\dfrac{\delta\Gamma_{\textrm{ind}}}{\delta \sigma}\right |
		=	\left\langle T\right\rangle,
\end{equation} 
where $|$ means the replacement $(\bar{g}_{\mu\nu},\bar{\rho},\bar{S}_\mu)\rightarrow \left(g_{\mu\nu},\rho, S_\mu\right)$ and $\sigma\rightarrow 0$. The general covariant solution for the anomaly-induced action consists of the
sum of the nonlocal part and local part with an arbitrary conformal invariant functional which plays the role of integration constant \cite{Buch-Shap(21)}. Following \cite{Shap et All (22),Gui-Shap(22)}, we know that the nonlocal part of the anomaly-induced action can provide the effective potential of the theory in the low energy regime. Consider the covariant nonlocal solution of equation (\ref{Tra-IndAc}),
	\begin{align}
		\Gamma_{\textrm{ind, nonloc}} & =\dfrac{b}{8}\int_{x}\int_{y}\left(E_{4}-\dfrac{2}{3}\square R\right)_{x}G\left(x,y\right)\left(E_{4}-\dfrac{2}{3}\square R\right)_{y}\nonumber \\
		& +\dfrac{1}{4}\int_{x}\int_{y}X\left(x\right)G\left(x,y\right)\left(E_{4}-\dfrac{2}{3}\square R\right)_{y},\label{eq:3.8}
	\end{align}
where we used the notation $\int_{x}\equiv\int d^{4}x\sqrt{-g\left(x\right)}\,$ and (\ref{eq:3.8-2}). The detail of the calculation can be found in \cite{Shap-Tors (02),Gui-Shap(22),Buch-Shap(21),Buchbinder(85),Ferreira (17)}.

Let us provide some commentary on the equation (\ref{eq:3.8}). Observe that to obtain the solution (\ref{eq:3.8}), we introduce the Green function
\begin{equation}
	\left(\sqrt{-g}\Delta_4\right)_x
		G\left(x,y\right)=\delta\left(x,y\right),
		\nn
\end{equation}
where $\Delta_4$ is the Paneitz operator \cite{Fradkin (84),S. Paneitz (08)},
\begin{equation} \label{TC-2}
	\Delta_4=\square^2
		+2R^{\mu\nu}\nabla_{\mu}\nabla_{\nu}
		-\dfrac{2}{3}R\square
		+\dfrac{1}{3}\nabla^\mu R\nabla_\mu\,,
\end{equation}
which is conformal invariant, i.e., $\sqrt{-g}\Delta_4=\sqrt{-\bar{g}}\bar{\Delta}_4$. An another important expressions is defined as
\begin{equation}
	X=wC^{2}+\beta_{1}\rho^{2}S^{2}-\dfrac{1}{4}\beta_{2}S_{\mu\nu}^{2}\,,\label{eq:3.8-2}
\end{equation}
because, such a term is conformal invariant. Note that the difference with the similar equation in \cite{Gui-Shap(22)} is the scalar term $\beta_1 \rho^2 S^2$, but this difference does not affect the general formalism to deriving effective potential from the anomaly-induced action. 

\section{Anomaly-induced effective action in the IR}
 As we already know that the nonlocal form of the effective action was recently shown to admit the description in the IR limit,  we must follow the procedure presented in \cite{Gui-Shap(22), Shap et All (22)}(see also  \cite{Mot-Vau (06),Gia-Mot (09)} for other detailed). In this way, let us define some assumptions:

Assuming a relatively weak gravitational field in the early Universe, the torsion terms in (\ref{eq:3.8-2}) dominate over the square of the Weyl tensor. Consequently, we have the following hierachy:
		$\left|\rho^2S^{2}\right|\gg\left|C^{2}\right|$ and $\left|S_{\mu\nu}^{2}\right|\gg\left|C^{2}\right|$. 
		
In general relativity, the IR limit implies a weak gravitational field. Thus,  we can describe the weak gravity through a small metric perturbation, i.e,,   $\left|\square{R}\right|\gg \left|R^{2}_{\ldots}\right|$ for all curvature tensor. In this way, the Green function in the induced action reduces to
\begin{equation}
	G=\Delta_{4}^{-1}=\left(\square^{2}+2R^{\mu\nu}\nabla_{\mu}\nabla_{\nu}-\dfrac{2}{3}R\square+\dfrac{1}{3}R^{\mu}\nabla_{\mu}\right)^{-1}
	\approx\dfrac{1}{\square{{}^2}}.
\end{equation}
Thus, after a small algebra, the nonlocal part of the effective action, equation (\ref{eq:3.8}),
becomes 
\begin{align}
	\Gamma_{\textrm{ind, nonloc}}^{\textrm{IR}} & =\dfrac{1}{6}\int_{x}\int_{y}\left(\beta_{1}\rho^{2}S^{2}+\dfrac{1}{4}\beta_{2}S_{\mu\nu}^{2}\right)_{x}\left(\dfrac{1}{\square}\right)_{x,y}R\left(y\right). \label{eq:4.0}
\end{align}

Here, we aim to find the low-energy effective action pertaining to the scalar-torsion sector mentioned in equation (\ref{eq:4.0}). To achieve this, we need to isolate the conformal factor from the metric. Subsequently, we draw upon an analogy with the renormalization group-based derivation of effective action \cite{Shap (92)}. Therefore, at one loop, 
let us consider conformal transformations involving (\ref{C-T-R})
together with
\begin{align} \label{C-T-R-2}
	\dfrac{1}{\square}=\dfrac{1}{\bar{\square}}e^{2\sigma}\quad\textrm{and}\quad R=e^{-2\sigma}\left[\bar{R}-6\bar{\square}\sigma+\mathcal{O}\left(\sigma^{2}\right)\right],
\end{align}
then later assume linear approximation in $\sigma$ terms in the previous equation. 
After integration with the delta function, the low-energy effective action of the scalar-torsion sector from (\ref{eq:4.0}) becomes
\begin{equation}
	\Gamma_{\textrm{ind, nonloc}}^{\textrm{IR}}=-\int_{x}\left(\beta_{1}\bar{\rho}^{2}\bar{S}^{2}+\dfrac{1}{4}\beta_{2}\bar{S}_{\mu\nu}^{2}\right)\sigma\left(x\right).\label{eq:4.2}
\end{equation}
We can see through this result that the anomaly-induced
action is a local version of the renormalization group corrected classical
action of torsion.

At this point, we will provide a concise overview of our classical action of torsion, found \cite{Shap-Tors (02),Shap (92)}, and  defined as 
\begin{equation}
	S_{\textrm{tor}}=-\int d{{}^4}x\sqrt{-g}\left\{ a_{1}\rho{{}^2}S^{2}
    +\dfrac{a_{2}}{4}S_{\mu\nu}^{2}
	\right\}
	+\textrm{surface terms}, 
	\label{eq:4.3}
\end{equation}
where we consider the parameters $a_1$ and $a_2$ as arbitrary variables. We impose the condition $a_{1}<0$ and $a_{2}>0$ to ensure that the energy is positively defined for propagating torsion, and to prevent the inclusion of tachyons in our theory due to the possible instability.

We shall come back to equation (\ref{eq:4.2}), which is related to loop correction to the classical action of torsion, and it will enable us to obtain effective action through the substitution process denoted as 
\begin{equation}
	a_{1}\rightarrow a_{1}-\beta_{1}\sigma\left(x\right),\quad a_{2}\rightarrow a_{2}-\beta_{2}\sigma\left(x\right),\label{eq:4.4}
\end{equation} 
in accordance with the renormalization group \cite{Shapiro (08),Buch-Shap(21),Buch(84)}, where the usual constant scaling parameter in curved spacetime is replaced by the local function $\sigma$.

Now, this approach can be employed to derive the one-loop effective potential. To do this, we make the identification 
\begin{equation}
	\sigma\rightarrow-\dfrac{1}{2}\ln\left(\dfrac{\rho^{2}+\eta^{2}S^{2}}{\mu^{2}}\right),\label{eq:4.5}
\end{equation}
that is consistent with the scalar and torsion parts of (\ref{C-T-R}). Here, $\mu$ represents the mass-dimension renormalization parameter, and the argument of the logarithm must remain dimensionless. Consequently, the quantity $\rho^2 + \eta^2S^2$ possesses a mass dimension of two.

Using (\ref{eq:4.4}) and (\ref{eq:4.5}) in the classical action (\ref{eq:4.3}),
we can easily get the effective action of torsion
\begin{align}
	\Gamma_{\textrm{tors}}  =&-\int d^{4}x\sqrt{-g}\left\{ \left[a_{1}+\dfrac{\beta_{1}}{2}\ln\left(\dfrac{\rho^{2}+\eta^{2}S^{2}}{\mu^{2}}\right)\right]\rho^{2}S^{2}\right.\nonumber \\
	& \left.+\dfrac{1}{4}\left[a_{2}+\dfrac{\beta_{2}}{2}\ln\left(\dfrac{\rho^{2}+\eta^{2}S^{2}}{\mu^{2}}\right)\right]S_{\mu\nu}^{2}+\ldots\right\}, \label{eq:4.6}
\end{align}
where surface terms have been neglected. The effective potential part is defined as a zero-order term in the derivative expansion of the effective action, review the equation (\ref{Geff}). Obviously
\begin{equation}
	V_{S^2}=\left[a_{1}+\dfrac{\beta_{1}}{2}\ln\left(\dfrac{\rho^{2}+\eta^{2}S^{2}}{\mu^{2}}\right)\right]\rho^{2}S^{2}.\label{eq:4.7}
\end{equation}
The torsion potential described above is similar to the one found at \cite{Gui-Shap(22)}, and both differ from the Coleman-Weinberg potential \cite{Coleman (73)}. Consequently, we conclude that spontaneous symmetry breaking does not occur in this theory. However, this conclusion does not extend to the effective potential of the NJL model, where the contributions from potentials in both flat and curved spacetimes must be considered, as discussed in the following section.

\section{Effective Potential and Gap Equation}

The calculation of the effective potential of the NJL model in a weakly varying gravitational background (which was obtained in terms of the Riemann normal coordinate expansion) with the pseudoscalar field set to $\pi=0$ can be seen in \cite{Marc,Eliz(94),Inag(93),Inagaki (97)}.
The effective potential in the flat spacetime  boils down to
\begin{equation}
	V_{\textrm{flat}}=V_{0}+\dfrac{1}{2\lambda}\rho^{2}-\dfrac{1}{\left(4\pi\right)^{2}}\left[\rho^{2}\Lambda^{2}+\Lambda^{4}\ln\left(1+\dfrac{\rho^{2}}{\Lambda^{2}}\right)-\rho^{4}\ln\left(1+\dfrac{\Lambda^{2}}{\rho^{2}}\right)\right], \label{eq:5.1}
\end{equation}
and the contribution of the curvature term to the effective potential has the form
\begin{equation}
	V_{R}=-\dfrac{1}{\left(4\pi\right)^{2}}\dfrac{R}{6}\left[-\rho^{2}\ln\left(1+\dfrac{\Lambda^{2}}{\rho^{2}}\right)+\dfrac{\Lambda^{2}\rho^{2}}{\Lambda^{2}+\rho^{2}}\right]. \label{eq:5.1-2}
\end{equation}
where $V_{0}=V\left.\right|_{\rho=0}$ and $\Lambda$ is
cut-off scale. However, we are interested in the expression of the
	effective potential in curved spacetime with external torsion. So let us consider (\ref{eq:5.1}) and (\ref{eq:5.1-2}) together with the potential
	(\ref{eq:4.7}), where we take the identification that $\mu^{2}\rightarrow\Lambda^{2}$, and we consider $N=1$ in (\ref{eq:3.7}) to ensure that the effective potential of torsion is indeed related to the NJL model, since the potential of the torsion was derived for $N$ fermions. Therefore, the expression
	for the effective potential in curved spacetime with torsion is given
	by
\begin{align} 
	V_{\textrm{eff}}&  =V_{0}+\dfrac{1}{2\lambda}\rho^{2}-\dfrac{1}{\left(4\pi\right)^{2}}\left[\rho^{2}\Lambda^{2}+\Lambda^{4}\ln\left(1+\dfrac{\rho^{2}}{\Lambda^{2}}\right)-\rho^{4}\ln\left(1+\dfrac{\Lambda^{2}}{\rho^{2}}\right)\right]\nonumber \\
	& -\dfrac{1}{\left(4\pi\right)^{2}}\dfrac{R}{6}\left[-\rho^{2}\ln\left(1+\dfrac{\Lambda^{2}}{\rho^{2}}\right)+\dfrac{\Lambda^{2}\rho^{2}}{\Lambda^{2}+\rho^{2}}\right]
	+\left[a_{1}+\dfrac{\beta_{1}}{2}\ln\left(\dfrac{\rho^{2}+\eta^{2}S^{2}}{\Lambda^{2}}\right)\right]\rho^{2}S^{2}.\label{eq:5.2}
\end{align}

To facilitate the study of the qualitative characteristics of the previous potential, for the sake of both generality and simplification in our computations, we have introduced the parameter $a_1= -\eta^2/\left(4\pi\right)^2$ and we take $\eta^2 S^2\rightarrow S^2$.

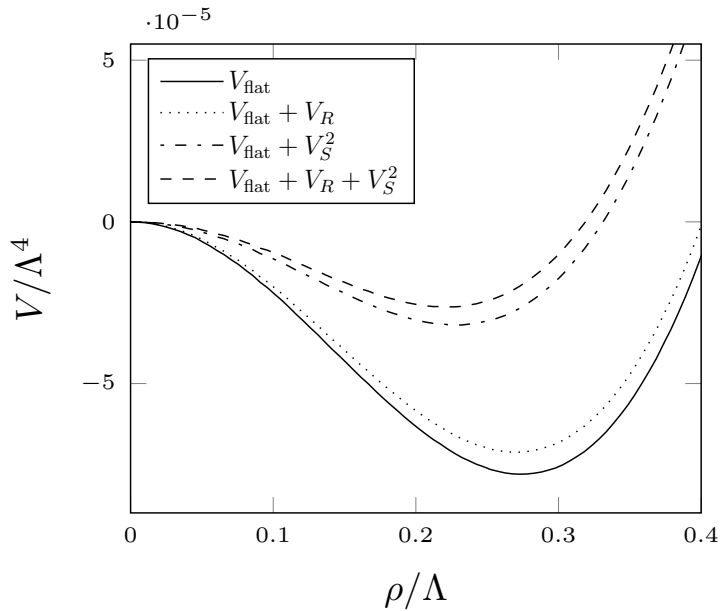
\begin{figure}[H]
	\centering
	\begin{tikzpicture}[scale=1.2]
		\begin{axis}[
			xmin = 0, xmax = 0.4,
			ymax=0.000055, ymin=-0.00009,
			xtick distance = 0.1,
			yticklabel style = {font=\tiny},
			xticklabel style = {font=\tiny},
			xlabel={\footnotesize{${\rho}/{\Lambda}$}} ,
			ylabel={\footnotesize{$V/{\Lambda^4}$}} ,
			legend cell align = {left},
			legend pos = north west,
			legend style={nodes={scale=0.6, transform shape}}, 
			]
			]
			\addplot[
			domain = 0:0.5 ,
			samples =100 ,
			color = black ,
			]
			{ (1/((4*pi)^2))*((3/5)*(x^2)-ln(1+x^2)+(x^4)*ln(1+1/(x^2))) };
			\addplot[
			domain = 0:0.5 ,
			samples =100 ,
			dotted,
			]
			{ (1/((4*pi)^2))*(((3/5)*(x^2)-ln(1+x^2)+(x^4)*ln(1+1/(x^2)))
				-(0.05/6)*(-(x^2)*ln(1+1/(x^2))+(x^2)/(1+x^2))) };
			\addplot[
			domain = 0:0.5 ,
			samples =100 ,
			dash pattern=on 1pt off 3pt on 3pt off 3pt,
			]
			{ (1/((4*pi)^2))*(((3/5)*(x^2)-ln(1+x^2)+(x^4)*ln(1+1/(x^2)))
				+(0.01)*(x^2)*(-1-(4)*(ln(0.01+x^2)))) };
			\addplot[
			domain = 0:0.5 ,
			samples =100 ,
			dashed,
			]
			{ (1/((4*pi)^2))*(((3/5)*(x^2)-ln(1+x^2)+(x^4)*ln(1+1/(x^2)))
				-(0.05/6)*(-(x^2)*ln(1+1/(x^2))+(x^2)/(1+x^2))
				+(0.01)*(x^2)*(-1-(4)*(ln(0.01+x^2)))) };
			
			\legend{
				$V_{\textrm{flat}}$, 
				$V_{\textrm{flat}}+V_R$,
				$V_{\textrm{flat}}+V_S^{2}$,
				$V_{\textrm{flat}}+V_R+V_S^{2}$
			}

		\end{axis}
	\end{tikzpicture}
	\caption{\textit{The behavior of the different effective potentials $V$ with $\lambda/\lambda_0=1.25$,
			curvature $R/\Lambda^{2} =0.05$ and torsion $S^{2}/\Lambda^{2}=0.01$.}}
	\label{fig 2}
\end{figure}

Fig \ref{fig 2} illustrates the behavior of effective potential $V_{\textrm{eff}}$ for different structural configurations, including the effective potential of the NJL model in flat spacetime, its implementation with curvature (see \cite{Eliz(94),Inag(93),Inagaki (97)} for more details), torsion contribution, and finally the effective potential with curvature and torsion. It is worth mentioning that the dynamical symmetry breaking in the NJL model in flat spacetime occurs when the coupling constant $\lambda$ is $\lambda>\lambda_{0}={4\pi^{2}}/{\Lambda^{2}}$.

	The spontaneous symmetry breaking in the effective potential of the NJL model differs from that in (\ref{eq:4.7}) due to the presence of contributions from both the flat spacetime and curvature. At this point, we can observe that in the presence of torsion and curvature, the local minimum of the effective potential becomes less pronounced, as is evident in the plot of the potential of the NJL model in Fig \ref{fig 2}. 

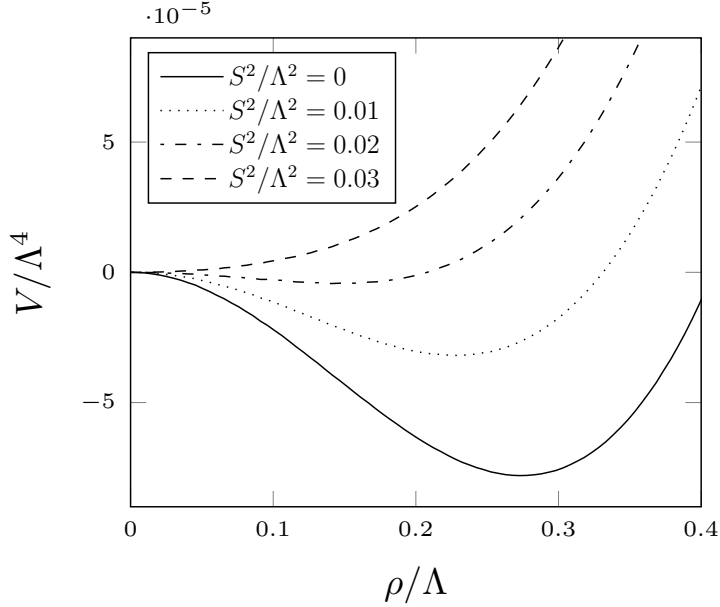
\begin{figure}[H]
	\centering
	\begin{tikzpicture}[scale=1.2]
		\begin{axis}[
			xmin = 0, xmax = 0.4,
			ymax=0.00009, ymin=-0.00009,
			yticklabel style = {font=\tiny},
			xticklabel style = {font=\tiny},
			xlabel={\footnotesize{${\rho}/{\Lambda}$}} ,
			ylabel={\footnotesize{$V/{\Lambda^4}$}} ,
			legend cell align = {left},
			legend pos = north west,
			legend style={nodes={scale=0.6, transform shape}}, 
			]
			]
			\addplot[
			domain = 0:0.5 ,
			samples =100 ,
			color = black ,
			]
			{ (1/((4*pi)^2))*((3/5)*(x^2)-ln(1+x^2)+(x^4)*ln(1+1/(x^2))) };
			\addplot[
			domain = 0:0.5 ,
			samples =100 ,
			dotted,
			]
			{ (1/((4*pi)^2))*(((3/5)*(x^2)-ln(1+x^2)+(x^4)*ln(1+1/(x^2)))
				+(0.01)*(x^2)*(-1-(4)*(ln(0.01+x^2))) };
			\addplot[
			domain = 0:0.5 ,
			samples =100 ,
			dash pattern=on 1pt off 3pt on 3pt off 3pt,
			]
			{ (1/((4*pi)^2))*(((3/5)*(x^2)-ln(1+x^2)+(x^4)*ln(1+1/(x^2)))
				+(0.02)*(x^2)*(-1-(4)*(ln(0.02+x^2)))) };
			\addplot[
			domain = 0:0.5 ,
			samples =100 ,
			dashed,
			]
			{ (1/((4*pi)^2))*(((3/5)*(x^2)-ln(1+x^2)+(x^4)*ln(1+1/(x^2)))
				+(0.03)*(x^2)*(-1-(4)*(ln(0.03+x^2))) };
			
			\legend{
				$S^2/\Lambda^2=0$, 
				$S^2/\Lambda^2=0.01$,
				$S^2/\Lambda^2=0.02$,
				$S^2/\Lambda^2=0.03$
			}

		\end{axis}
	\end{tikzpicture}
	\caption{\textit{The behavior of the effective potential $V$ with $\lambda/\lambda_0=1.25$ as a function of different torsion values}}
	\label{fig 2-2}
\end{figure}

The behavior of the effective potential of the NJL model as a function of different values of the torsion term can be observed in the qualitative graph shown in Fig \ref{fig 2-2}. Notably, the effective potential does not have a local minimum when the value reaches $S^2/\La^2\geq0.03$, demonstrating that the possibility of dynamic symmetry breaking is not achieved for these values of the square of torsion.

To explicitly observe spontaneous symmetry breaking in the NJL model, we begin by examining the gap equation related to the effective potential (\ref{eq:5.2}). This equation represents the vacuum expectation value, giving us the dynamical mass of the fermion. In our case, $\rho_0$ plays the role of the dynamical mass of the fermion, i.e., $\rho_0=M$. For more information about the equation gap in flat spacetime, see, e.g., \cite{Klev(92),Buballa(05)}, and in curved spacetime, see, e.g., \cite{Eliz(94),Inag(93),Inagaki (97)}. 
\begin{align}
	\left.\dfrac{\partial V\left(\rho,0\right)}{\partial\rho}\right|_{\rho=\rho_{0}}  = &\,\,\dfrac{\rho_{0}}{\lambda}-\dfrac{\rho_{0}}{4\pi^{2}}\left[\Lambda^{2}-\rho_{0}^{2}\ln\left(1+\dfrac{\Lambda^{2}}{\rho_{0}^{2}}\right)\right]\nonumber\\
	& -\dfrac{\rho_{0}}{48\pi^{2}}R\left[-\ln\left(1+\dfrac{\Lambda^{2}}{\rho_{0}^{2}}\right)+\dfrac{\Lambda^{2}}{\Lambda^{2}+\rho_{0}^{2}}+\dfrac{\Lambda^{4}}{\left(\Lambda^{2}+\rho_{0}^{2}\right)^{2}}\right]\nonumber\\
	& +2\left[a_{1}+\dfrac{\beta_{1}}{2}\ln\left(\dfrac{\rho_{0}^{2}+\eta^{2}S^{2}}{\Lambda^{2}}\right)\right]\rho_{0} S^{2}
	+\dfrac{\beta_{1}\rho_{0}^{3}S^{2}}{\eta^{2}S^{2}+\rho_{0}^{2}}=0.
	\label{eq:5.3}
\end{align}

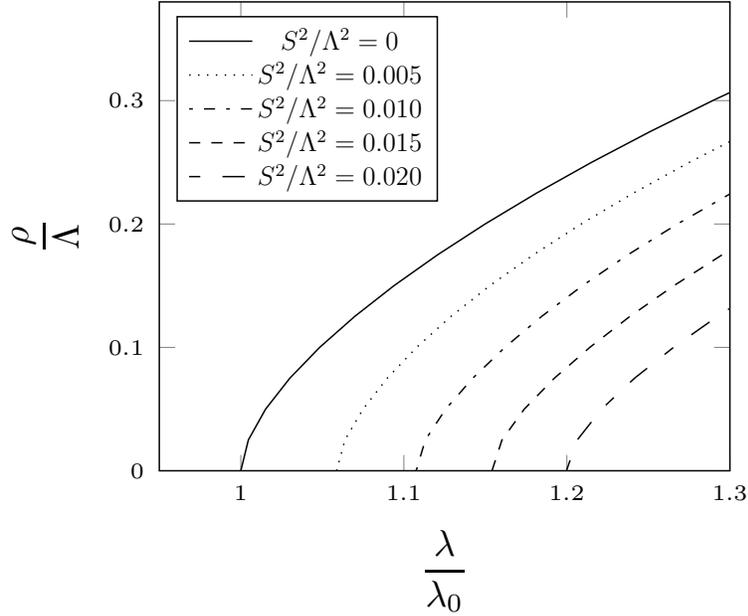
\begin{figure}[H]
	\centering
	\begin{tikzpicture}[scale=1.2]
		\begin{axis}[
			xlabel={\footnotesize$\dfrac{\lambda}{\lambda_{0}}$},
			ylabel={\footnotesize$\dfrac{\rho}{\Lambda}$},
			xmin=0.95, xmax=1.3,
			ymin=0, ymax=0.38,
			yticklabel style = {font=\tiny},
			xticklabel style = {font=\tiny},
			legend pos=north west,
			legend style={nodes={scale=0.6, transform shape}}
			]
			\addplot[
			samples=100,
			]
			coordinates {
				
				(1, 0)
				(1.004632854, 0.025)
				(1.015212867, 0.05)
				(1.030048638, 0.075)
				(1.048384194, 0.1)
				(1.069775922, 0.125)
				(1.093937553, 0.15)
				(1.120675059, 0.175)
				(1.149853326, 0.2)
				(1.181377032, 0.225)
				(1.215178799, 0.25)
				(1.251211427, 0.275)
				(1.289442594, 0.3)
				(1.329851113, 0.325)
				(1.372424215, 0.35)
				(1.417155538, 0.375)
				(1.464043605, 0.4)
				(1.513090668, 0.425)
				(1.564301803, 0.45)
				(1.617684207, 0.475)
				(1.673246648, 0.5)
				(1.730999023, 0.525)
				(1.790952013, 0.55)
				(1.853116806, 0.575)
				(1.917504879, 0.6)
				(1.984127822, 0.625)
				(2.052997205, 0.65)
				(2.124124467, 0.675)
				(2.197520837, 0.7)
				(2.27319727, 0.725)
				(2.351164397, 0.75)
				(2.431432497, 0.775)
				(2.514011469, 0.8)
				(2.598910817, 0.825)
				(2.686139647, 0.85)
				(2.775706661, 0.875)
				(2.867620158, 0.9)
				(2.961888045, 0.925)
				(3.05851784, 0.95)
				(3.157516683, 0.975)
				(3.258891353, 1)
			};
			
			
			\addplot[
			samples=100,
			dotted,
			]
			coordinates {(1.053167224, 0)
				(1.055749621, 0.025)
				(1.061661392, 0.05)
				(1.07160852, 0.075)
				(1.085754626, 0.1)
				(1.103705927, 0.125)
				(1.125017468, 0.15)
				(1.149334491, 0.175)
				(1.176395715, 0.2)
				(1.206011172, 0.225)
				(1.238041834, 0.25)
				(1.272384719, 0.275)
				(1.30896254, 0.3)
				(1.347716508, 0.325)
				(1.388601262, 0.35)
				(1.431581236, 0.375)
				(1.476627999, 0.4)
				(1.523718276, 0.425)
				(1.572832456, 0.45)
				(1.62395345, 0.475)
				(1.677065799, 0.5)
				(1.732154992, 0.525)
				(1.789206917, 0.55)
				(1.848207436, 0.575)
				(1.909142046, 0.6)
				(1.971995611, 0.625)
				(2.036752155, 0.65)
				(2.103394694, 0.675)
				(2.171905113, 0.7)
				(2.242264068, 0.725)
				(2.31445092, 0.75)
				(2.388443679, 0.775)
				(2.464218983, 0.8)
				(2.54175207, 0.825)
				(2.621016784, 0.85)
				(2.701985577, 0.875)
				(2.784629524, 0.9)
				(2.86891835, 0.925)
				(2.954820455, 0.95)
				(3.042302954, 0.975)
				(3.131331713, 1)
				
			};
			\addplot[
			samples=100,
			dash pattern=on 1pt off 3pt on 3pt off 3pt,
			]
			coordinates {
			(1.095414315, 0)
			(1.098087706, 0.025)
			(1.103296628, 0.05)
			(1.11129598, 0.075)
			(1.122828028, 0.1)
			(1.138062215, 0.125)
			(1.156811847, 0.15)
			(1.178788218, 0.175)
			(1.203714915, 0.2)
			(1.231360452, 0.225)
			(1.261539664, 0.25)
			(1.294106194, 0.275)
			(1.328943919, 0.3)
			(1.365959576, 0.325)
			(1.405076944, 0.35)
			(1.446232391, 0.375)
			(1.489371493, 0.4)
			(1.534446471, 0.425)
			(1.581414235, 0.45)
			(1.63023488, 0.475)
			(1.680870532, 0.5)
			(1.73328445, 0.525)
			(1.787440327, 0.55)
			(1.843301755, 0.575)
			(1.900831807, 0.6)
			(1.959992714, 0.625)
			(2.020745634, 0.65)
			(2.083050468, 0.675)
			(2.146865744, 0.7)
			(2.21214853, 0.725)
			(2.278854401, 0.75)
			(2.346937415, 0.775)
			(2.416350133, 0.8)
			(2.487043651, 0.825)
			(2.558967651, 0.85)
			(2.632070466, 0.875)
			(2.706299164, 0.9)
			(2.781599637, 0.925)
			(2.8579167, 0.95)
			(2.935194195, 0.975)
			(3.01337511, 1)
				
			};
			
			
			\addplot[
			samples=100,
			dashed,
			]
			coordinates {
				(1.134418563, 0)
				(1.137239805, 0.025)
				(1.142290333, 0.05)
				(1.14932988, 0.075)
				(1.159175134, 0.1)
				(1.172328595, 0.125)
				(1.18888177, 0.15)
				(1.20869684, 0.175)
				(1.231554914, 0.2)
				(1.257231095, 0.225)
				(1.285523193, 0.25)
				(1.316258538, 0.275)
				(1.349292148, 0.3)
				(1.384502235, 0.325)
				(1.42178542, 0.35)
				(1.461052503, 0.375)
				(1.502224967, 0.4)
				(1.545232183, 0.425)
				(1.590009213, 0.45)
				(1.636495084, 0.475)
				(1.68463145, 0.5)
				(1.734361559, 0.525)
				(1.78562945, 0.55)
				(1.838379341, 0.575)
				(1.892555171, 0.6)
				(1.948100262, 0.625)
				(2.004957076, 0.65)
				(2.063067055, 0.675)
				(2.12237053, 0.7)
				(2.182806674, 0.725)
				(2.244313512, 0.75)
				(2.30682796, 0.775)
				(2.370285895, 0.8)
				(2.434622255, 0.825)
				(2.499771154, 0.85)
				(2.565666016, 0.875)
				(2.632239718, 0.9)
				(2.699424753, 0.925)
				(2.767153387, 0.95)
				(2.835357826, 0.975)
				(2.903970391, 1)	
			};
			
			
			\addplot[
			dash pattern=on 3pt off 6pt on 6pt off 6pt,
			samples=100,
			]
			coordinates {
				(1.171619972, 0)
				(1.174604242, 0.025)
				(1.179676613, 0.05)
				(1.186181991, 0.075)
				(1.194857602, 0.1)
				(1.206373387, 0.125)
				(1.22103792, 0.15)
				(1.238874205, 0.175)
				(1.259753045, 0.2)
				(1.283486912, 0.225)
				(1.309879647, 0.25)
				(1.338747971, 0.275)
				(1.36992836, 0.3)
				(1.403277239, 0.325)
				(1.438668587, 0.35)
				(1.475990818, 0.375)
				(1.515143748, 0.4)
				(1.556035942, 0.425)
				(1.598582508, 0.45)
				(1.642703313, 0.475)
				(1.688321568, 0.5)
				(1.73536272, 0.525)
				(1.783753601, 0.55)
				(1.833421783, 0.575)
				(1.884295105, 0.6)
				(1.936301333, 0.625)
				(1.989367937, 0.65)
				(2.043421958, 0.675)
				(2.098389952, 0.7)
				(2.154197988, 0.725)
				(2.210771707, 0.75)
				(2.268036411, 0.775)
				(2.32591719, 0.8)
				(2.384339069, 0.825)
				(2.443227179, 0.85)
				(2.502506939, 0.875)
				(2.562104245, 0.9)
				(2.621945666, 0.925)
				(2.681958643, 0.95)
				(2.742071686, 0.975)
				(2.802214561, 1)		
			};

			\legend{
				$S^2/\Lambda^2=0$, 
				$S^2/\Lambda^2=0.005$,
				$S^2/\Lambda^2=0.010$,
				$S^2/\Lambda^2=0.015$,
				$S^2/\Lambda^2=0.020$
			}				
		\end{axis}
	\end{tikzpicture}
	\caption{\textit{The plot of dynamical mass as a function of different torsion values of $S^{2}/\Lambda^{2}$.}}
	\label{fig 3}
\end{figure}

\begin{figure}[H]
	\centering
	\begin{tikzpicture}[scale=1.2]
		\begin{axis}[
			xlabel={\footnotesize$\dfrac{\lambda_0}{\lambda}$},
			ylabel={\footnotesize$\dfrac{S^2}{\Lambda^2}$},
			xmin=0.2, xmax=1.2,
			ymin=0, ymax=0.3,
			yticklabel style = {font=\tiny},
			xticklabel style = {font=\tiny},
			xtick distance = 0.2,
			xtick={0,0.2,0.40,0.6,0.8,1.0,1.2},
			ytick={0,0.10,0.20,0.30,0.40},
			]
			
			\addplot[
			samples=100,
			]
			coordinates {
				
			(0.999986184, 0)
			(0.94950701, 0.005)
			(0.91288678, 0.01)
			(0.881499032, 0.015)
			(0.853509264, 0.02)
			(0.828046212, 0.025)
			};
			\addplot[
			samples=100,
			dashed,
			]
			coordinates {		
			(0.804596711, 0.03)
			(0.782821679, 0.035)
			(0.762480118, 0.04)
			(0.743391833, 0.045)
			(0.725416957, 0.05)
			(0.708443754, 0.055)
			(0.692380898, 0.06)
			(0.677152343, 0.065)
			(0.662693779, 0.07)
			(0.64895011, 0.075)
			(0.635873601, 0.08)
			(0.623422501, 0.085)
			(0.611559975, 0.09)
			(0.600253291, 0.095)
			(0.589473166, 0.1)
			(0.579193249, 0.105)
			(0.569389704, 0.11)
			(0.56004086, 0.115)
			(0.551126936, 0.12)
			(0.542629799, 0.125)
			(0.534532769, 0.13)
			(0.526820449, 0.135)
			(0.519478585, 0.14)
			(0.512493939, 0.145)
			(0.505854189, 0.15)
			(0.499547834, 0.155)
			(0.493564116, 0.16)
			(0.487892949, 0.165)
			(0.482524858, 0.17)
			(0.477450928, 0.175)
			(0.47266275, 0.18)
			(0.468152387, 0.185)
			(0.463912326, 0.19)
			(0.459935454, 0.195)
			(0.456215019, 0.2)
			(0.452744612, 0.205)
			(0.44951813, 0.21)
			(0.446529767, 0.215)
			(0.443773982, 0.22)
			(0.44124549, 0.225)
			(0.438939238, 0.23)
			(0.436850395, 0.235)
			(0.434974334, 0.24)
			(0.433306621, 0.245)
			(0.431843004, 0.25)
			(0.4305794, 0.255)
			(0.429511888, 0.26)
			(0.428636694, 0.265)
			(0.427950192, 0.27)
			(0.427448885, 0.275)
			(0.427129406, 0.28)
			(0.426988508, 0.285)
			(0.427023058, 0.29)
			(0.42723003, 0.295)
			(0.427606502, 0.3)
			(0.428149648, 0.305)
			(0.428856736, 0.31)
			(0.429725121, 0.315)
			(0.430752243, 0.32)
			(0.431935622, 0.325)
			(0.433272852, 0.33)
			(0.434761604, 0.335)
			(0.436399615, 0.34)
			(0.43818469, 0.345)
			(0.440114697, 0.35)
			(0.442187567, 0.355)
			(0.444401286, 0.36)
			(0.446753899, 0.365)
			(0.449243502, 0.37)
			(0.451868245, 0.375)
			(0.454626325, 0.38)
			(0.457515987, 0.385)
			(0.460535523, 0.39)
			(0.463683268, 0.395)
			(0.466957599, 0.4)
			(0.470356933, 0.405)
			(0.473879727, 0.41)
			(0.477524475, 0.415)
			(0.481289708, 0.42)
			(0.485173991, 0.425)
			(0.489175924, 0.43)
			(0.493294139, 0.435)
			(0.497527299, 0.44)
			(0.501874097, 0.445)
			(0.506333258, 0.45)	
			};
			legend
			\node[] at (axis cs: 0.42,0.07) {\scriptsize{Broken Phase}};
			\node[] at (axis cs: 0.801,0.2) {\scriptsize{Symmetric Phase}};
			\node[] at (axis cs: 0.828, 0.025) {\scriptsize{$\bullet$}};
		\end{axis}
	\end{tikzpicture}
	
	\caption{\textit{The torsion is represented as a function of the coupling constant that demarcates the boundary between the region of the symmetric phase and one of the broken phase.}} 
	\label{fig 4}
\end{figure}
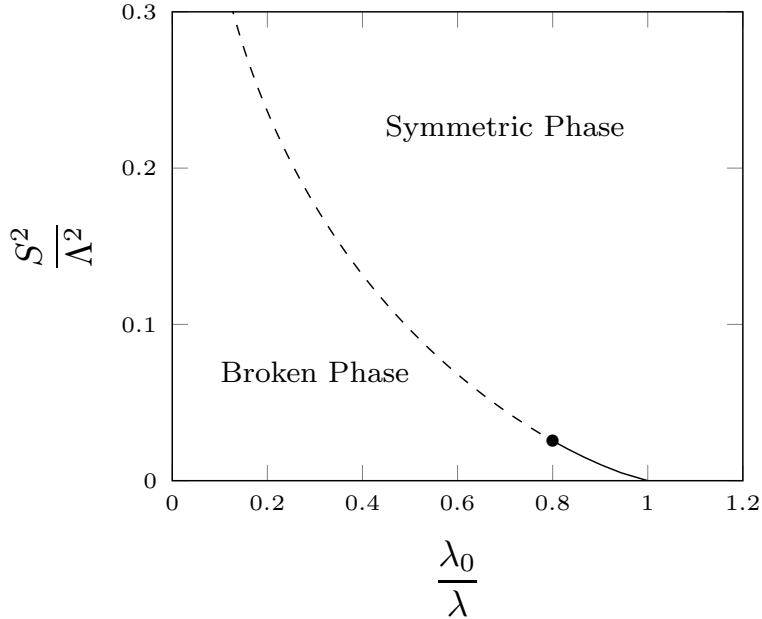

The relationship between dynamical mass of the fermion and the coupling constant for $R=0$ can be seen in Fig \ref{fig 3} for different values of torsion. In this graph,  one notes that the coupling constant varies concerning the torsion term, where $\lambda$ increases together with the term of the torsion $S^{2}/\Lambda^2$. 
In this way, we can conclude that the scalar-torsion sector influences the mass generation of fermions. A detailed analysis of the torsion field as a function of the coupling constant becomes evident  in Fig \ref{fig 4}.

The graph in Fig \ref{fig 4} consists of two regions, one with a symmetric phase and another with a broken phase. The critical curve shown in the plot is a function of the torsion with respect to the coupling constant. Through  Fig \ref{fig 4}, we can make some observations. First, only the lower part (continuous line) has physical relevance that corresponds to $\lambda_0/\lambda\leq0.8$. For the rest (dashed line) the potential does not have broken phase, since, for torsion values with $S^2/\Lambda^{2}\geq0.03$ there is not a local minimum in the effective potential of the NJL model. 

Additionally, even when $\lambda$ remains within the region of the broken phase in the NJL model in the flat spacetime, the chiral symmetry can be restored for positive values of $S^{2}/\Lambda^{2}$ above the critical curve. 

Finally, we investigate the influence of the parameter $ a_1 $ on the effective potential. To achieve this, we rewrite $ a_1 $ as $ n\, a_1 $ in equation (\ref{eq:5.2}), where $ n $ belongs to the set of real numbers. This allows us to generate the graph in Fig \ref{fig 5}. In this graph, we observe that the contribution of $ n\, a_1 $ for $ n > 1 $ in the torsion sector favors the local minimum of the effective potential, which becomes more pronounced, favoring the occurrence of spontaneous symmetry breaking in the model.
\begin{figure}[H]	
	\centering
	\begin{tikzpicture}[scale=1.2]
		\begin{axis}[
			xmin = 0, xmax = 0.45,
			ymax=0.00006, ymin=-0.0001,
			xtick distance = 0.1,
			yticklabel style = {font=\tiny},
			xticklabel style = {font=\tiny},
			xlabel={\footnotesize{${\rho}/{\Lambda}$}} ,
			ylabel={\footnotesize{$V/{\Lambda^4}$}} ,
			legend cell align = {left},
			legend pos = north west,
			legend style={nodes={scale=0.6, transform shape}}, 
			]
			]
			\addplot[
			domain = 0:0.5 ,
			samples =100 ,
			color = black ,
			]
			{ (1/((4*pi)^2))*((3/5)*(x^2)-ln(1+x^2)+(x^4)*ln(1+1/(x^2))) };
			\addplot[
			domain = 0:0.5 ,
			samples =100 ,
			dotted,
			]
			{ (1/((4*pi)^2))*(((3/5)*(x^2)-ln(1+x^2)+(x^4)*ln(1+1/(x^2)))
				+(0.01)*(x^2)*(-1-(4)*(ln(0.01+x^2)))) };
			\addplot[
			domain = 0:0.5 ,
			samples =100 ,
			dash pattern=on 1pt off 3pt on 3pt off 3pt,
			]
			{ (1/((4*pi)^2))*(((3/5)*(x^2)-ln(1+x^2)+(x^4)*ln(1+1/(x^2)))
				+(0.01)*(x^2)*(-4-(4)*(ln(0.01+x^2)))) };
				\addplot[
			domain = 0:0.5 ,
			samples =100 ,
			dashed,
			]
			{ (1/((4*pi)^2))*(((3/5)*(x^2)-ln(1+x^2)+(x^4)*ln(1+1/(x^2)))
				+(0.01)*(x^2)*(-6-(4)*(ln(0.01+x^2)))) };
			
			\legend{
				$V_{\textrm{flat}}$,
				$V_{\textrm{flat}}+V_S^{2}\textrm{,}$ $n=1$, 
				$V_{\textrm{flat}}+V_S^{2}\textrm{,}$ $n=4$,
				$V_{\textrm{flat}}+V_S^{2}\textrm{,}$ $n=6$
			}

		\end{axis}
	\end{tikzpicture}
	\caption{\textit{The effective potentials $V$ with $\lambda/\lambda_0=1.25$
	 and torsion $S^{2}/\Lambda^{2}=0.01$ for different values of 
			$n$.}}
	\label{fig 5}
\end{figure}

\section{Conclusions and discussions}
The effective action of the Nambu-Jona-Lasinio in an external gravitational field with torsion in the $1/N$ expansion was found by introducing the auxiliary fields, i.e., using the technique of bosonization. The effective potential of the torsion is calculated using the new technique based on the nonlocal part of anomaly-induced action which was recently found to produce the effective potential in the low-energy limit.

Also, we were able to identify a Coleman-Weinberg potential for the NJL model with some specific values for the external torsion field, see Fig. \ref{fig 2-2} and for certain values of $a_1$ in equation (\ref{eq:5.2}), see Fig. \ref{fig 5}. Furthermore, we concluded that the scalar-torsion sector has a more significant influence than curvature on the effective potential well, as can be seen in Fig. \ref{fig 2}. Later on, we computed the gap equation (\ref{eq:5.3}) with the intention of explicitly observing spontaneous symmetry breaking. 

We concluded that the first-order phase transition occurs when the torsion value changes, i.e., the coupling constant has its value changed as a function of the torsion, as can be seen in Fig \ref{fig 3}. Looking along the graph in Fig \ref{fig 4}, we conclude that only the lower part (continuous line) holds physical relevance. Our decision is based on the observation that higher values of torsion prevent the effective potential of the NJL model from exhibiting behavior similar to that of the Coleman-Weinberg potential. Additionally, one notices that the chiral symmetry can be restored even if the coupling constant is kept in the broken phase region, but only for values of torsion above the critical curve contained in Fig \ref{fig 4}.

In the recent paper \cite{Gus-Ant (23)}, chiral symmetry breaking was investigated independently in the bosonized version of the Nambu-Jona-Lasino model in a Riemann-Cartan background. In principle, both studies can agree on the influence of the torsion field on the effective potential of the NJL model when considering $ n > 1 $, as shown in Fig \ref{fig 5}. It is worth noting that the effective potential of the scalar-torsion sector (\ref{eq:4.7}) may not be suitable for describing dynamical symmetry breaking. This characteristic is attributed to the negative value of $ \beta_1 $, as demonstrated in \cite{Gui-Shap(22)} for a similar theory. On the other hand, as suggested in \cite{Gui-Shap(22)}, the function $ \beta_1 $ could assume a change of sign at higher loops, thus reconciling the results of both works completely.

\section{Acknowledgments}
The author would like to thank I. L. Shapiro for comments and encouragement in the course of
this work. I am also grateful to CAPES for supporting my Ph.D. project.

\section{Data Availability Statement}

 No Data associated in the manuscript.

\end{document}